\documentclass[reprint,aps,ams,amsmath,prx,twocolumn,longbibliography]{revtex4-1}
\usepackage{graphics}
\usepackage{graphicx}
\usepackage{epsfig}
\usepackage{amsmath}
\usepackage{amssymb}
\usepackage{amsfonts}
\usepackage{bm}
\usepackage{color}
\usepackage{bbm}
\usepackage[bookmarks=false,linkcolor=blue,urlcolor=blue,colorlinks,citecolor=blue]{hyperref}
\usepackage[normalem]{ulem}
\usepackage{comment}
\usepackage{soul}
\usepackage{libertine}
\DeclareMathOperator{\St}{St}

\begin{document}

\title{Transport signatures of plasmon fluctuations in electron hydrodynamics}

\author{Dmitry Zverevich}
\affiliation{Department of Physics, University of Wisconsin-Madison, Madison, Wisconsin 53706, USA}

\author{Alex Levchenko}
\affiliation{Department of Physics, University of Wisconsin-Madison, Madison, Wisconsin 53706, USA}

\date{November 4, 2023}

\begin{abstract}
In two-dimensional electron systems, plasmons are gapless and long-lived collective excitations of propagating charge density oscillations. 
We study the fluctuation mechanism of plasmon-assisted transport in the regime of electron hydrodynamics.  
We consider pristine electron liquids where charge fluctuations are thermally induced by viscous stresses and intrinsic currents, 
while attenuation of plasmons is determined by the Maxwell mechanism of charge relaxation. We show that while the contribution of
plasmons to the shear viscosity and thermal conductivity of a Fermi liquid is small, plasmon resonances in the bilayer devices enhance  
the drag resistance. In systems without Galilean invariance, fluctuation-driven contributions to dissipative coefficients can be described only in 
terms of hydrodynamic quantities: intrinsic conductivity, viscosity, and plasmon dispersion relation.
\end{abstract}

\maketitle

\section{Introduction}

Hydrodynamic effects in electron transport can occur in solids at intermediate temperatures when the system is sufficiently pure, see reviews \cite{NGMS,Lucas-Fong,ALJS,Narozhny,Scaffidi} for the detailed discussion of recent results. Indeed, Gurzhi  \cite{Gurzhi} argued early on that if electron-electron interactions provide the most frequent scattering mechanism so that the corresponding mean free path is the shortest length scale in the problem, then one could use approximate local conservation laws to develop an effective hydrodynamic description. The hydrodynamic equations can be obtained by expanding the equations of motion for the element of the fluid in gradients of the velocity and thermodynamic quantities up to terms of second order in the spatial derivatives \cite{LL-V6}. Alternatively, these equations can be derived from a more microscopic Boltzmann kinetic theory by projecting collision terms into the slow modes \cite{LL-V10}. In this principal approximation, the structural form of hydrodynamic equations follows uniquely from the general conservation laws of particle, momentum, energy densities, and underlying symmetry of the system.    

When going beyond the leading approximation one can consider several different types of corrections to hydrodynamics. First are the usual gas-kinetic corrections obtained by Burnett \cite{Burnett} based on Boltzmann equation with an extension of the Chapman-Enskog expansion method. These corrections lead to the appearance of the higher-order gradient terms. This approach can be made systematic in dilute systems, e.g. gases, however, it is uncontrolled for fluids where there is no small parameter. Second, are the correlation effects 
that can arise in electron systems subjected to the long-range disorder potential such as strongly-correlated high-mobility semiconductor devices \cite{Spivak,AKS} and graphene devices \cite{Lucas,Li}. It can be shown that in these systems hydrodynamic equations retain their principal form on the scales large as compared to the disorder correlation radius but with the renormalized quantities and dissipative coefficients. Third is thermal fluctuation corrections in pristine systems due to the presence of long-lived collective modes, particularly acoustic fluctuations. It was shown by Andreev \cite{Andreev} that the fluctuation mechanism is always the basic one at sufficiently small gradients. In neutral normal fluids long-wavelength thermal fluctuations result in nonanalytical corrections thus leading to essentially nonlocal equations. Interestingly, these corrections contain no new parameters and can be expressed solely in terms of the thermodynamic functions and dissipative coefficients that enter the hydrodynamic equations in the main approximation. 

In electron liquids, the relevant collective modes are propagating charge density fluctuations --- plasmons \cite{Pines}. In two-dimensional electron systems, plasmons are low-lying gapless and long-lived excitations \cite{Fetter,Ando}. While plasmons were meticulously studied over the years, most recently in the context of graphene and surface states of topological insulators \cite{Vafek,Barlas,DasSarma,Hwang,MacDonald,Principi,Vignale,Levitov,DasSarmaLi,Fogler,LucasDasSarma,HwangDasSarma,Svintsov,Polini,Lewandowski,Titov,Oriekhov}, the role of plasmons in electron hydrodynamic behavior was not systematically addressed. In the hydrodynamic limit, plasmons can be thermally excited by fluctuating viscous stresses and intrinsic currents. In bilayer devices, one finds both acoustic and optical plasmons branches. In systems without Galilean invariance, these modes are attenuated by the Maxwell mechanism of charge relaxation. In the Galilean invariant case, when intrinsic conductivity vanishes, the decay of plasmons is governed by viscous diffusion.  Our estimations show that fluctuation corrections of propagating plasmons to viscosity and thermal conductivity of a Fermi liquid are small. We further identify examples of transport phenomena where plasmons play a dominant role in the hydrodynamic regime. Building on our recent work on the near-field energy transfer \cite{Zverevich} we consider Coulomb drag resistance in the electronic double-layers.   

The presentation is organized as follows. In Sec. \ref{sec:hydro} we summarize the main ingredients of the theory of hydrodynamic fluctuations in electron liquids that form the basis of our analysis. This presentation parallels earlier works \cite{Apostolov,Levchenko} on the technically overlapping topics. In Sec. \ref{sec:drag} we apply this formalism to determine the nonequilibrium dynamical structure factor of an electron fluid to the linear order in the hydrodynamic velocity. This result enables the calculation of the drag resistance near plasmon resonances in electron double-layer devices. In Sec. \eqref{sec:plasmons} we introduce the kinetic equation for plasmons in the relaxation time approximation and apply it to estimate the contribution of plasmons to the viscosity and thermal conductivity of electron fluid. As any real device is prone to some degree of disorder in Sec. \ref{sec:disorder} we recall the impact of impurities and long-range density inhomogeneities on the plasmon broadening. 

\section{Hydrodynamic fluctuations}\label{sec:hydro}

To have a self-contained presentation we provide a brief account of the hydrodynamic theory with an inclusion of stochastic Langevin forces. 
For normal fluids this formalism was developed by Landau and Lifshitz \cite{LL}, and generalized by Khalatnikov \cite{Khalatnikov} for the case of superfluids. 
The pedagogical presentation of lectures on hydrodynamic fluctuations can be found in Refs. \cite{Forster,Kovtun}. 
In applications to electron liquids in quantum materials one has to add the Coulomb law as appropriate for the charged system, 
and incorporate additional terms in transport currents for the systems that are generically not Galilean invariant. 

For this purpose and having in mind applications to interactively-coupled transport in electron double-layered devices, we consider a planar geometry of two conducting two-dimensional sheets separated by the distance $d$. Applicability of the hydrodynamic approximation requires us to consider transport regime when the intralayer electron mean-free path $l$ is short as compared to the interlayer spacing, $l\ll d$. In each layer the thermally-driven spatial and temporal fluctuations of the particle density $\delta n(\bm{r},t)$ render the corresponding current fluctuations $\delta\bm{j}_n(\bm{r},t)$ that follow each other in accordance with the continuity equation \cite{LL-V6}
\begin{equation}\label{eq:continuity-n}
\partial_t\delta n+\bm{\nabla}\cdot\delta\bm{j}_n=0.
\end{equation}
Fluctuations of the particle current density comprise of several contributions \footnote{Note that we define electrical current fluctuations as $\delta\bm{j}_e=e\delta\bm{j}_n$.}
\begin{equation}\label{eq:djn}
\delta\bm{j}_n=\bm{v}\delta n+n\delta\bm{v}+\frac{\sigma}{e^2}\delta\bm{F}+\delta\bm{\xi}. 
\end{equation}
The first term in the above expression is the convective part of fluctuations in the presence of macroscopic hydrodynamic flow of the fluid with the velocity $\bm{v}(\bm{r},t)$. 
The second term describes fluctuations in the hydrodynamic velocity. The third term captures current fluctuations generated by fluctuating electromotive force $\delta\bm{F}$. This term is present in systems with broken Galilean invariance, which have nonvanishing intrinsic conductivity $\sigma$. The last term is dictated by the fluctuation-dissipation theorem and describes random Langevin currents whose correlation function at temperature $T$ is given by \cite{Kogan}
\begin{equation}\label{eq:j-j}
\langle\delta\xi_i(\bm{r},t)\delta\xi_j(\bm{r}',t')\rangle=2T\frac{\sigma}{e^2}\delta_{ij}\delta(\bm{r}-\bm{r}')\delta(t-t'), 
\end{equation}
where $\langle\ldots\rangle$ denotes thermal average \footnote{In this work use the natural units and set Planck's and Boltzmann's constants to unity $\hbar=k_B=1$. Therefore, temperature has units of energy and square of electron charge has units of velocity.}. In principle, particle current fluctuations in Eq. \eqref{eq:djn} may also include thermoelectric contributions generated by the temperature fluctuations, e.g. $\alpha\bm{\nabla}\delta T$, where $\alpha$ is an intrinsic thermoconductivity.  However, for the type of effects that we consider below, these terms lead to insignificant corrections, which we thus neglect.   

Thermal fluctuations also lead to the entropy density fluctuations of the fluid $\delta s(\bm{r},t)$ and associated with it fluctuations of the respective entropy current density 
$\delta\bm{j}_s(\bm{r},t)$. These quantities are also linked by the continuity equation
\begin{equation}\label{eq:continuity-s}
\partial_t\delta s+\bm{\nabla}\cdot\delta\bm{j}_s=0.
\end{equation}
Without thermoelectric effects the entropy current density also has four contributions similar to Eq. \eqref{eq:djn}
\begin{equation}\label{eq:djs}
\delta\bm{j}_s=\bm{v}\delta s+s\delta\bm{v}-\frac{\kappa}{T}\bm{\nabla}\delta T+\frac{\delta\bm{\zeta}}{T}. 
\end{equation}
Here the first two terms are completely analogous to that in Eq. \eqref{eq:djn}. The third term describes entropy fluxes due to finite thermal conductivity $\kappa$. The last term is the associated Langevin thermal noise whose correlation function is given by  
\begin{equation}
\langle\delta\zeta_i(\bm{r},t)\delta\zeta_j(\bm{r}',t')\rangle=2\kappa T^2\delta_{ij}\delta(\bm{r}-\bm{r}')\delta(t-t').  
\end{equation}

For a fluid with the mass density $\rho$ the evolution of the momentum density $\bm{p}=\rho\bm{v}$ is governed by the Navier-Stokes equation 
\begin{equation}\label{eq:NS}
\rho(\partial_t+\bm{v}\cdot\bm{\nabla})\delta\bm{v}=-\bm{\nabla}\delta\hat{\Pi}-en\bm{\nabla}\delta\Phi, 
\end{equation}
which we present here in its linearized form with respect to fluctuations. The first term on the right-hand-side of Eq. \eqref{eq:NS} captures fluctuations of the momentum flux tensor 
\begin{equation}
\delta\Pi_{ij}=\delta P\delta_{ij}-\delta\Sigma_{ij}.
\end{equation} 
It includes local hydrodynamics fluctuations in the pressure of a fluid 
\begin{equation}
\delta P=\left(\frac{\partial P}{\partial n}\right)_s\delta n+\left(\frac{\partial P}{\partial n}\right)_v\delta s,
\end{equation}
that couples Eq. \eqref{eq:NS} to the continuity equations \eqref{eq:continuity-n} and \eqref{eq:continuity-s}, and fluctuations of viscous stresses 
\begin{equation}\label{eq:dSigma}
\delta\Sigma_{ij}=\eta(\partial_i\delta v_j+\partial_j\delta v_i-\delta_{ij}\partial_k\delta v_k)+\delta\Xi_{ij}, 
\end{equation}
that are expressed via gradients of the velocity field, where $\eta$ is the shear viscosity. The random viscous Langevin sources in Eq. \eqref{eq:dSigma} are described by the correlation function of the form \cite{LL}
\begin{align}\label{eq:Xi-Xi}
&\langle\delta\Xi_{ik}(\bm{r},t)\delta\Xi_{lm}(\bm{r}',t')=\nonumber \\ 
&2\eta T(\delta_{il}\delta_{km}+\delta_{im}\delta_{kl}-\delta_{ik}\delta_{lm})\delta(\bm{r}-\bm{r}')\delta(t-t'). 
\end{align}
For simplicity, we neglected terms with the bulk viscosity. As is known, bulk viscosity vanishes in the systems with quadratic and linear dispersion relations \cite{LL-V10}. 
Finally, the second term on the right-hand-side of Eq. \eqref{eq:NS} describes the flow of momentum due to fluctuations of the long-range Coulomb interaction. The corresponding fluctuations of the electric potential $\delta\Phi$ are related to the electron density fluctuation $\delta n$ by the Poisson equation. It should be noted that in the geometry of a bilayer $\delta\Phi$ includes both potential due to density fluctuations in a given layer, as well as dynamically screened potential arising from the density fluctuations in the other layer.  

The formalism of hydrodynamic theory of fluctuations allows computation of various correlation functions in any concrete setup. The approach to a given problem of interest is rather straightforward and can be summarized as follows. One considers Langevin sources $\delta\bm{\xi}, \delta\bm{\zeta}, \delta\Xi_{ij}$ as given functions fluctuating in space and time. The linearized equations of motion then can be solved for $\delta n, \delta\bm{v}, \delta P$ with an account of the proper boundary conditions. As a result these quantities are exprressed as linear functionals of the source fields. Therefore, any quadratic form with respect to $\delta n, \delta\bm{v}, \delta P$ can be expressed via quadratic average of sources with the help of the fluctuation-dissipation relations. Upon thermal averaging $\langle\ldots\rangle$ the auxiliary sources drop out and the result is expressed via a handful number of dissipative coefficients and thermodynamic quantities of the system.      

In this work we calculate the dynamic structure factor of the fluid which is formally defined as the density-density correlation function. This object is finite even in equilibrium, it carries information about the collective modes in the system and obeys rather generic properties such as Kramer-Kronig relations and sum rules \cite{Forster,LL-V9}. Furthermore, being interested in the transport effects we take one step further, and calculate correction to the structure factor to the linear order in hydrodynamic velocity. These results enable applications to nonlocal transport effects in bilayers such as drag friction.    

\section{Plasmon-enhanced Coulomb drag resistivity}\label{sec:drag}

Coulomb drag \cite{RMP-Drag} is the useful experimental technique to directly probe the strength of electronic correlations that can be quantified via measured drag resistance. 
The setup consists of two spatially-separated and electrically-isolated conducting layers, where one (active) layer is driven out of equilibrium and the resulting nonlocal response is measured in the other (passive) layer that can be dragged along since electrons interact via the long-range Coulomb potential. Importantly for our applications, Coulomb drag was recently measured in both monolayer and bilayer graphene double-layers \cite{Tutuc,Geim,Kim-Tutuc,Dean,Kim}. To the best of our knowledge, the existing calculations of this effect \cite{HwangDrag,Katsnelson,Carrega,Ostrovsky,Lux1,Schutt,Song,Lux2} didn't address the role of plasmons in the hydrordynamic regime. Thus far such analysis was carried out only in the Galilean invariant systems \cite{Apostolov,Chen}. Here we provide solution to this problem with generalizations as appropriate for systems with broken Galilean invariance.     

\subsection{Linear response analysis}

For simplicity we consider symmetrical layers with average carrier density $n$. We work in the limit $k_Fd\gg1$, where $k_F$ is the Fermi momentum. We denote density fluctuations in each layer as $\delta n_{1,2}(\bm{r},t)$ and respective fluctuations of velocity as $\delta\bm{v}_{1,2}(\bm{r},t)$. In the active layer we impose finite (in average) homogeneous hydrodynamic velocity $\bm{v}$. For all fluctuating quantities, including Langevin sources, we introduce Fourier components $\{\delta n, \delta\bm{v}, \delta\Phi\}\propto \exp(-i\omega t+i\bm{qr})$. In these notations the linearized continuity equation \eqref{eq:continuity-n} takes the form 
\begin{align}\label{eq:continuity-linear}
-i\omega\delta n_1+in(\bm{q}\cdot\delta\bm{v}_1)+i(\bm{q}\cdot\bm{v})\delta n_1\nonumber \\ 
+\frac{\sigma}{e^2}q^2e\delta\Phi_1+i(\bm{q}\cdot\delta\bm{\xi}_1)=0.
\end{align}
In the fluctuating electromotive force, Coulomb potential includes both fluctuations of the density in the layer-1 as well as screened fluctuations of the density in the layer-2, namely  
\begin{equation}
\delta\Phi_1(\bm{q},\omega)=\frac{2\pi e}{\epsilon q}\left[\delta n_1(\bm{q},\omega)+e^{-qd}\delta n_2(\bm{q},\omega)\right],
\end{equation}
where $\epsilon$ is the dielectric constant of the material surrounding the electron layers. The continuity equation for the density fluctuations $\delta n_2$ in the passive layer is the same, one only has to interchange indices $1\leftrightarrow2$ and take $\bm{v}\to0$. 
The linearized Navier-Stokes equation \eqref{eq:NS} in the active layer takes the form 
\begin{align}\label{eq:NS-linear}
-i\rho\omega\delta\bm{v}_1+i\rho(\bm{q}\cdot\bm{v})\delta\bm{v}_1=-ien\bm{q}\delta\Phi_1+i(\bm{q}\cdot\delta\bm{\Sigma}_1).
\end{align}
Here we made one approximation by neglecting pressure fluctuations $\delta P$ as compared to fluctuations in the Coulomb potential $\delta\Phi$. 
This is legitimate as the long-range nature of Coulomb interaction dominates for fluctuations in the long wave length limit when $q\to0$.    
We also note that with this approximation, the entropy continuity equation decouples and thus entropy fluctuations will not contribute the to drag. 
Previous analysis \cite{Apostolov} showed that density-density coupling induced by thermal expansion of the fluid and temperature fluctuations leads to a very small drag that can be disregarded. It is convenient to multiply Eq. \eqref{eq:NS-linear} by one extra power of $\bm{q}$ to have equation in the scalar form. We then notice that $\bm{q}\cdot(\bm{q}\cdot\delta\bm{\Sigma})=i\eta q^2(\bm{q}\cdot\delta\bm{v})+\bm{q}\cdot(\bm{q}\cdot\delta\bm{\Xi})$. As a next step, we rewrite Eqs. \eqref{eq:continuity-linear} and \eqref{eq:NS-linear}
equivalently 
\begin{subequations}
\begin{align}
n(\bm{q}\cdot\delta\bm{v}_1)-\omega\delta n_1-i\gamma_q(\delta n_1+e^{-qd}\delta n_2)=\nonumber \\ 
-(\bm{q}\cdot\bm{v})\delta n_1-(\bm{q}\cdot\delta\bm{\xi}_1),
\end{align}
\begin{align}
(\omega+i\omega_\eta)n(\bm{q}\cdot\delta\bm{v}_1)-\omega^2_p(\delta n_1+e^{-qd}\delta n_2)=\nonumber \\ 
n(\bm{q}\cdot\bm{v})(\bm{q}\cdot\delta\bm{v}_1)-\frac{n}{\rho}\bm{q}\cdot(\bm{q}\cdot\delta\bm{\Xi}_1),
\end{align}
\end{subequations}
where we introduced 
\begin{equation}\label{eq:energies}
\omega_q=\sqrt{\frac{2\pi(ne)^2q}{\rho\epsilon}}, \quad \gamma_q=\frac{2\pi\sigma q}{\epsilon},\quad \omega_\eta=\frac{\eta q^2}{\rho}. 
\end{equation}
There are two more equations of the same kind for the passive layer. At this point, one can exclude fluctuations of velocity to arrive at the coupled equations that govern density fluctuations only. For this purpose, it is convenient to symmetrize density fluctuations and introduce $\delta n_\pm=\delta n_1\pm\delta n_2$, and similarly for all otgher quantities e.g. $\delta\bm{\xi}_\pm=\delta\bm{\xi}_1\pm\delta\bm{\xi}_2$. In this basis of normal modes fluctuations $\delta n_\pm$ decouple. Finally, we introduce $\delta n_\pm=\delta n^{(0)}_\pm+\delta n^{(1)}_\pm+\ldots$, where $\delta n^{(0)}_\pm$ denote the equilibrium fluctuations induced by the Langevin sources, and $\delta n^{(1)}_\pm$ capture the nonequilibrium  advection of fluctuations by the hydrodynamic flow to the linear order in $\bm{v}$. After some algebra we find 
\begin{subequations}\label{eq:dn}
\begin{equation}
\delta n^{(0)}_\pm=-\frac{n}{\rho\Gamma_\pm}\bm{q}\cdot(\bm{q}\cdot\delta\bm{\Xi}_\pm)+\frac{\omega+i\omega_\eta}{\Gamma_\pm}(\bm{q}\cdot\delta\bm{\xi}_\pm),
\end{equation}
\begin{equation}
\delta n^{(1)}_\pm=\frac{(\bm{q}\cdot\bm{v})}{2\Gamma_\pm}\sum_{\pm}\left[\Upsilon_\pm\delta n^{(0)}_{\pm}-(\bm{q}\cdot\delta\bm{\xi}_\pm)\right].
\end{equation}
\end{subequations} 
Here we introduced 
\begin{subequations}
\begin{equation}
\Gamma_\pm(q,\omega)=\omega^2-\omega^2_\pm-\omega_\eta\gamma_\pm+i\omega(\omega_\eta+\gamma_\pm),
\end{equation}
\begin{equation}
\Upsilon_\pm(q,\omega)=2\omega+i(\omega_\eta+\gamma_\pm), 
\end{equation}
\end{subequations}
where 
\begin{equation}\label{eq:plasmons}
\omega^2_\pm=\omega^2_q(1\pm e^{-qd}),\quad \gamma_\pm=\gamma_q(1\pm e^{-qd}).
\end{equation}
To close the system of equations we need thermal averages which follow from Eqs. \eqref{eq:j-j} and \eqref{eq:Xi-Xi}
\begin{subequations}
\begin{align}
\langle (\bm{q}\cdot\delta\bm{\xi}_\pm)(\bm{q}\cdot\delta\bm{\xi}_\pm)\rangle=4Tq^2\sigma/e^2, \\ 
\langle \bm{q}\cdot(\bm{q}\cdot\delta\bm{\Xi}_\pm)\bm{q}\cdot(\bm{q}\cdot\delta\bm{\Xi}_\pm)\rangle=4Tq^4\eta.
\end{align}
\end{subequations}
It is clear from the obtained solution that the dynamical structure factor $\langle\delta n_\pm\delta n_\pm\rangle$ contains resonant denominators for certain values of frequencies and wave numbers of fluctuations. We thus first discuss roots of $\Gamma_\pm(q,\omega)$.  

\subsection{Hydrodynamic plasmon modes}

As is well known from the hydrodynamic linear response theory \cite{Forster}, zeros of $\Gamma_\pm(q,\omega)$ define complex frequencies of collective modes propagating in the system. The real part of frequency defines the dispersion relation of the mode, while the complex part defines its decay rate. The double-layer system supports two modes termed as
optical plasmon (OP) and acoustic plasmon (AP) \cite{Ando,Fetter}. This distinction holds for $qd\ll1$, while in the opposite limit dispersions smoothly connect to the plasmon frequencies of each individual layers. The OP corresponds to the in-phase oscillations of the electron density and its dispersion relation is similar to the plasmon frequency of a single layer with the square-root dependence on $q$, $\omega_+\propto\sqrt{q}$. The AP mode corresponds to the out-of-phase charge neutral oscillations and thus exhibits linear dispersion $\omega_-\propto q$.  

Attenuation of plasmons occurs differently in systems with or without Galilean invariance. Indeed, in the Galilean invariant system, $\sigma\to0$, the imaginary part of the root of $\Gamma_\pm$ depends only on $\omega_\eta$. Therefore, relaxation occurs via viscous diffusion and the corresponding rate is the same for both OP and AP that scales quadratically with the wave vector, namely $\Im \omega\propto q^2$. The fact that kinematic viscosity determines plasmon decay was discussed earlier in the context of 
conductivity of the classical two-dimensional electron gas \cite{Hruska}. If Galilean invariance is broken, the relaxation is dominated by the Maxwell mechanism of charge dissipation. 
This feature is important and was discussed in the context of nonlocal optical conductivity in graphene \cite{Briskot}. In a double-layer, the decay rate of OP is linear in $q$ whereas it scales quadratically for the AP. We see that regardless of the attenuation mechanism, both branches of charge oscillations are underdamped so that plasmons are well-defined and long-lived excitations in the hydrodynamic regime.  We note that plasmons remain underdamped in the high-frequency (kinetic) regime, $\omega\gg v_Fq$, where plasmon attenuation is dominated by the decay into two part particle-hole pairs. From the results of Refs. \cite{Glazman,Maslov} we can deduce that the corresponding rates are the same for both Galilean invariant and Dirac systems, modulo logarithmic factors, and scale as $\Im \omega\propto q^2$. Disorder leads to additional mechanism of plasmon broadening that we discuss separately in Sec. \ref{sec:plasmons}. 

\subsection{Drag force and resistivity}

We apply the above formalism to the Coulomb drag problem. The steady current in the active layer exerts the drag force on the passive layer. 
We relate the potential to density fluctuations by using the Poisson equation and thus express the drag force in terms of the density-density correlation function
\begin{equation}\label{eq:FD}
\bm{\mathcal{F}}_D=\int\frac{d\omega d^2q}{(2\pi)^3}(-i\bm{q})\frac{2\pi e^2}{\epsilon q}e^{-qd}
\langle\delta n_1(\bm{q},\omega)\delta n_2(-\bm{q},-\omega)\rangle.
\end{equation}
Knowing the drag force one readily finds the drag resistivity
\begin{equation} 
r_D = \bm{\mathcal{F}}_D/(e^2n^2\bm{v}).
\end{equation}
We observe that since viscous stresses do not correlate with intrinsic current fluctuations the respective contributions to the drag force 
can be considered separately. We focus on the latter terms as they yield the dominant contribution.  

\begin{figure}[t!]
  \centering
  \includegraphics[width=3.5in]{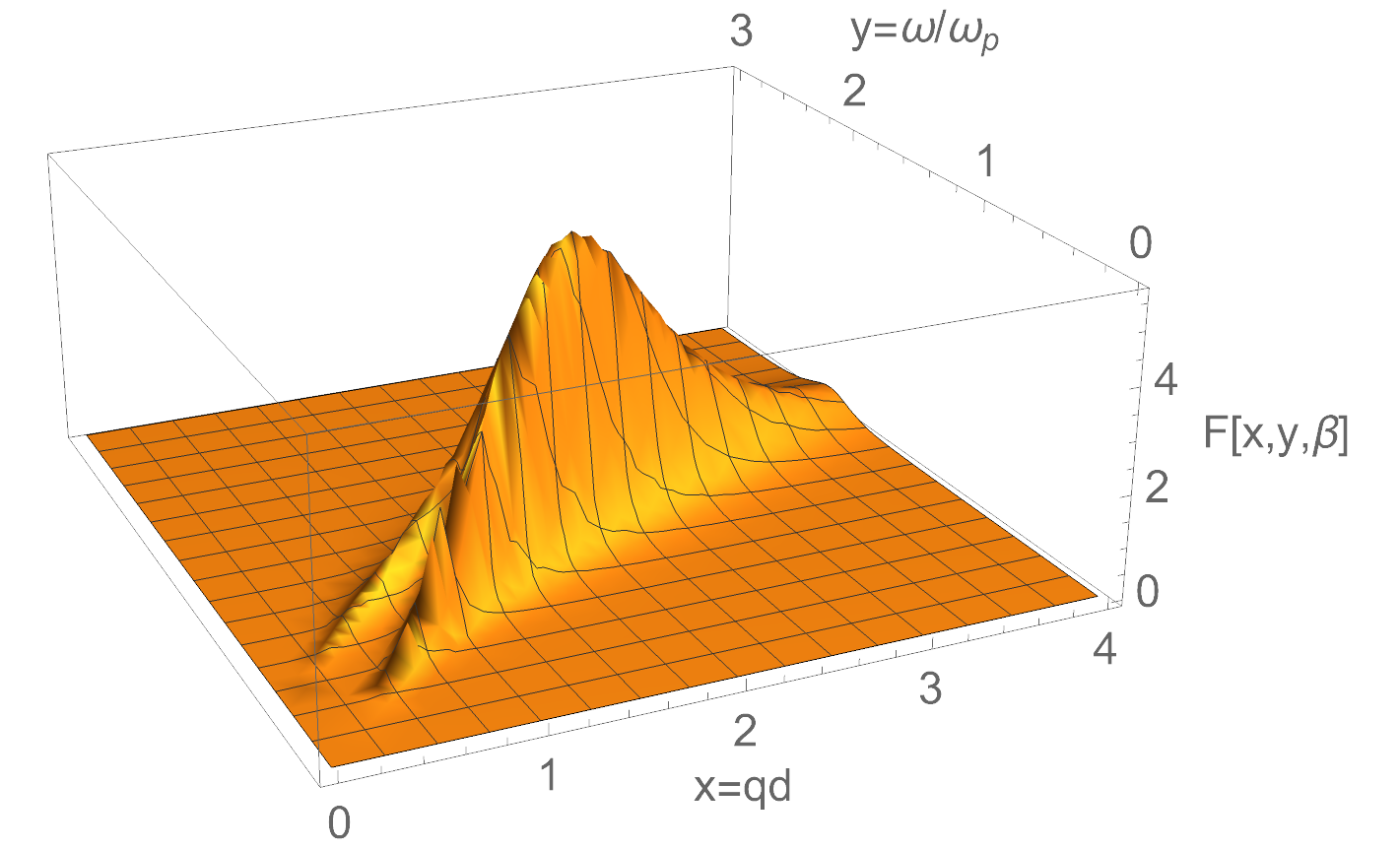}
  \caption{Dimensionless function $F(x,y,\beta)$ defined by Eq. \eqref{eq:F} plotted for $\beta=0.15$. The traces of plasmon resonances can be seen for momenta $x\gtrsim1$ as their spectral weight is suppressed at $x\to0$ thus making them invisible. The dimensionless parameter $\beta$ controls the broadening of plasmon branches and determines the maximum of $F$.} 
  \label{Fig-F3D}
\end{figure}

\begin{figure}[t!]
  \centering
  \includegraphics[width=3.25in]{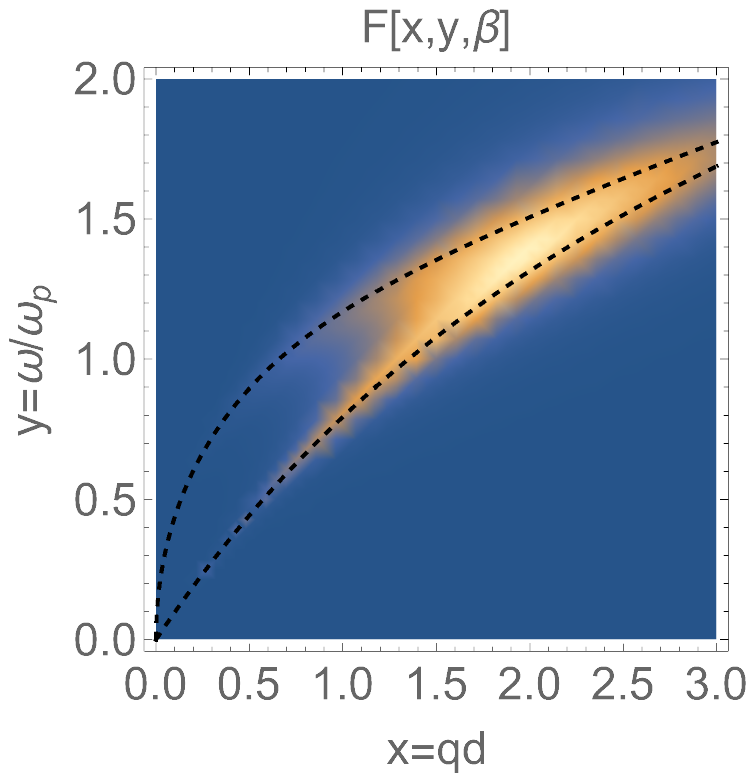}
  \caption{Dispersion laws $\omega=\omega_\pm(q)$ for OP and AP shown by dashed lines superimposed on top of the color plot that defines magnitude of the dimensionless function from Fig. \ref{Fig-F3D} plotted for $\beta=0.15$. The light area where OP and AP tend to merge corresponds to the maximum of $F\sim 4$.}
  \label{Fig-Plasmons}
\end{figure}

\begin{figure}[t!]
  \centering
  \includegraphics[width=3in]{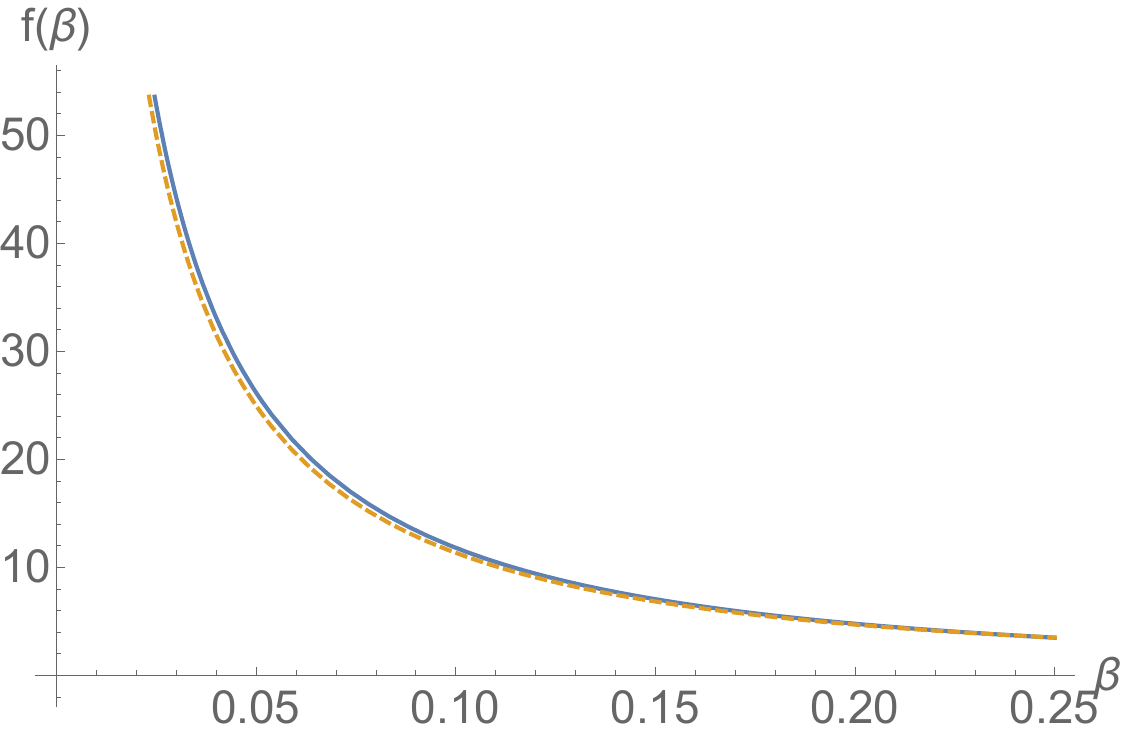}
  \caption{Plot of the dimensionless function $f(\beta)$ introduced in Eq. \eqref{eq:rD}. The solid line represents the result of a numerical integration, and the dashed line corresponds to the approximate analytical formula given in Eq. \eqref{eq:f-approx} and applicable for $\beta<1$. }
  \label{Fig-f}
\end{figure}

To calculate the average $\langle\delta n_1\delta n_2\rangle$ we transform it into the symmetrized basis. We notice that equilibrium parts of the average $\langle\delta n^{(0)}_\pm\delta n^{(0)}_\pm\rangle$ do not contribute to the force as they are isotropic thus average to zero upon $\bm{q}$ integration. To the linear order in $\bm{v}$   
we have two types of terms to consider: the same parity $\langle\delta n^{(0)}_\pm\delta n^{(1)}_\pm\rangle$ and the mixed parity $\langle\delta n^{(0)}_\pm\delta n^{(1)}_\mp\rangle$  
contributions. Using Eq. \eqref{eq:dn} we notice that both $\Gamma_\pm$ and $\Upsilon_\pm$ depend only on the absolute value of the wave vector $q=|\bm{q}|$, therefore they are even function with respect to $\bm{q}\to-\bm{q}$ exchange. In contrast, changing the sign of frequency, $\omega\to-\omega$ we get $\Gamma_\pm(q,-\omega)=\Gamma^*_\pm(q,\omega)$ and $\Upsilon_\pm(q,-\omega)=-\Upsilon^*_\pm(q,\omega)$. These properties result in the correlator  $\langle\delta n^{(0)}_\pm\delta n^{(1)}_\pm\rangle$ being frequency odd while correlator $\langle\delta n^{(0)}_\pm\delta n^{(1)}_\mp\rangle$ being frequency even. Thus, the former drops out upon the frequency integration in Eq. \eqref{eq:FD}. The remaining contributions come in the complex conjugated pairs. To simplify the formulas at the intermediate steps we notice that $\omega^2_+\gamma_--\omega^2_-\gamma_+=0$, which gives $\Im(\Pi_+\Pi^*_-)\approx\omega^3(\gamma_+-\gamma_-)$ and $\Im(\Upsilon^*_-\Gamma_++\Upsilon_+\Gamma^*_-)\approx3\omega^3(\gamma_+-\gamma_-)$.  
The approximate sign means that we neglected the shift of the plasmon pole $\propto \omega_\eta\gamma_\pm$ in the expressions for $\Gamma_\pm$, and we also neglected viscous damping of plasmons as compared to $\gamma_\pm$ in the imaginary part of $\Gamma_\pm$ and $\Upsilon_\pm$. Collecting all the pieces together we arrive at the following expression  
\begin{equation}
r_D=\frac{T\sigma}{2e^4n^2}\int\frac{d\omega d^2q}{(2\pi)^3}\left(\frac{2\pi e^2}{\epsilon q}\right)\frac{\omega^4q^4(\gamma_+-\gamma_-)e^{-qd}}{|\Gamma_+|^2|\Gamma_-|^2}.
\end{equation}
To deal with final integrations it is useful to rescale all frequencies $\omega$ in units of the plasma frequency $\omega_q$ taken at $q=1/d$, and also to rescale all momenta $q$ in units of $1/d$. The integral then depends on a single dimensionless parameter $\beta=(\gamma_q/\omega_q)_{q=1/d}$. These steps lead us to the final result for the drag resistance  
\begin{equation}\label{eq:rD}
r_D=\frac{\sigma}{4\pi^2e^4}\left(\frac{T}{E_F}\right)\left(\frac{1}{nd^2}\right)^2f(\beta).
\end{equation}
The dimensionless function $f(\beta)$ is obtained from the double-integral of the following function 
\begin{equation}\label{eq:F}
F(x,y,\beta)=\prod_{\pm}\frac{\sqrt{\beta}x^{5/2}y^2e^{-x}}{(y^2-x(1\pm e^{-x}))^2+\beta^2y^2x^2(1\pm e^{-x})^2}
\end{equation}
where $x=qd$ and $y=\omega/\omega_{q=1/d}$. In order to highlight importance of plasmons we plot $F(x,y,\beta)$ in Fig. \eqref{Fig-F3D}. Physically this function is defined by the product of the dynamical structure factor, the phase space volume, and the strength of Langevin fluxes. We see that the resonant contribution of plasmons to the drag resistance occurs for $q\sim 1/d$ and $\omega\sim\omega_{q=1/d}$, or equivalently in dimensionless notations $x\sim y\sim1$. For $(x,y)\ll1$ the spectral weight of plasmon resonances is suppressed and they are not clearly visible on the plot. On Fig. \ref{Fig-Plasmons} we further plot dispersions of optical and acoustic plasmons superimposed on top of the color plot of the function $F(x,y,\beta)$.  
We estimate the dimensionless parameter $\beta$ to be of the order 
\begin{equation}
\beta\sim \frac{\sigma}{e^2}\sqrt{\frac{e^2}{\epsilon v_F}}\frac{1}{\sqrt{k_Fd}}.
\end{equation}
Since we work in the limit $k_Fd\gg1$ it is small. In the limit $\beta\ll1$ we are able to extract an asymptotic expression for $f(\beta)$ in the form
\begin{equation}\label{eq:f-approx}
f(\beta)\approx\frac{\pi}{10} \ln^3(\Lambda)\ln\left(\frac{4}{\beta\ln\frac{\Lambda}{\beta}}\right),\quad \Lambda=\frac{2}{\beta\sqrt{\ln\frac{1}{\beta}}}.
\end{equation}
To assess the accuracy of this approximate formula we compare it to the result of the numerical integration and find an excellent agreement, see Fig. \ref{Fig-f}.    

In a similar way one can compute viscous contribution to the drag resistivity from Eqs. \eqref{eq:dn} and \eqref{eq:FD}. The final result can be found in the form  
\begin{equation}\label{eq:rD-visc}
r_D=\frac{\epsilon v_F}{e^4}\left(\frac{T}{E_F}\right)\frac{\eta/n}{(k_Fd)^5}g(\beta),
\end{equation}
where $g(\beta)$ is another dimensionless function that has logarithmic dependance on $\beta$ in the limit $\beta\ll1$. There are two key differences between $r_D$ given by Eqs. \eqref{eq:rD} and \eqref{eq:rD-visc}. First is that intrinsic mechanism is parametrically stronger as it decays as $1/d^4$ whereas viscous contribution diminishes much faster as $1/d^5$ at large interlayer separations. Therefore, drag effect is stronger in systems with broken Galilean invariance. This feature is qualitatively consistent with experimental observations. The second difference concerns the temperature dependence. In the Fermi liquid regime $\eta\propto 1/T^2$ therefore viscous contribution scales as $r_D\propto 1/T$ in the hydrodynamic limit. In contrast, for systems with broken Galilean invariance the temperature dependence of $r_D$ is determined by the product of intrinsic conductivity $\sigma(T)$ and an extra power of $T$ coming from the Langevin sources, modulo additional logarithmic term in $f(\beta)$. For example, in the monolayer and bilayer graphene, $\sigma(T)$ is known to be very weakly temperature dependent, therefore $r_D$ is approximately linear in temperature, $r_D\propto T$. It is striking to observe that at temperatures where $l\sim d$ the hydrodynamic result for drag resistance is parametrically larger than the conventional Fermi liquid result for the collisionless regime \cite{Smith,Kamenev,Flensberg}. This implies that frequent intralayer collisions strongly enhance Coulomb drag. We note that Eq. \eqref{eq:rD-visc} applies to the Galilean invariant systems except that $\beta$ should be replaced by a different dimensionless parameter that depends on viscosity $\beta\to\eta/(\rho d^2\omega_{q})$ with $q=1/d$ \cite{Apostolov}. Perhaps even more importantly, Eqs. \eqref{eq:rD} and \eqref{eq:rD-visc} apply to non-Fermi liquids as long as hydrodynamic limit can be justified \cite{Patel}.  

\section{Plasmon corrections to kinetic coefficients in Fermi liquids}\label{sec:plasmons}

As mentioned in the introduction, the fluctuational corrections to the viscosity of a two-dimensional neutral liquid diverge \cite{Andreev}, 
which means that hydrodynamic equations are nonlocal in two dimensions. On the other hand, the calculations presented in the preceding sections were based on the local form of the theory and assumed that the viscosity and thermal conductivity of the electron system are well-defined quantities. Thus it is instructive to estimate fluctuational corrections to the dissipative hydrodynamic coefficients of the two-dimensional charged electron liquid. We present a straightforward generalization of the procedure outlined in Ref. \cite{Hruska} to the case of systems with broken Galilean invariance. For this purpose it is convenient to introduce a distribution function of plasmon fluctuations $N_{\bm{q}}$ that obeys the kinetic equation
\begin{equation}\label{eq:KinEq}
\frac{\partial N_{\bm{q}}}{\partial t}+\frac{\partial\omega_q}{\partial\bm{q}}\bm{\nabla}N_{\bm{q}}=\St\{N_{\bm{q}}\}. 
\end{equation}
The microscopic derivation of the collision integral $\St\{N_{\bm{q}}\}$ that describes scattering of plasmons is a challenging task \cite{Fritz}. For the sake of estimation it suffice to use the relaxation-time-approximation for the linearized collision integral, $\St\{N_{\bm{q}}\}=-\gamma_q\delta N_{\bm{q}}$, where $\delta N_{\bm{q}}$ denotes the nonequilibrium part of the distribution and the intrinsic relaxation rate is given by Eq. \eqref{eq:energies}. 

The energy flux of plasmons can be written in the usual kinetic form
\begin{equation}
\bm{j}_\varepsilon=\int\frac{d^2q}{(2\pi)^2}\bm{v}_q\omega_qN_{\bm{q}},
\end{equation}
where $\bm{v}_q=\partial\omega_q/\partial\bm{q}$ is the group velocity of plasmons. Solving the kinetic equation to the linear order in a small temperature gradient $\bm{\nabla}T$ gives correction to the thermal conductivity $\bm{j}_\varepsilon=-\delta\kappa\bm{\nabla}T$,
\begin{equation}\label{eq:dkappa}
\delta\kappa=\frac{1}{2}\int\omega_q\tau_q\bm{v}^2_q\frac{\partial N_q}{\partial T}\frac{d^2q}{(2\pi)^2},
\end{equation}
where $\tau_q=1/\gamma_q$ and $N_q$ should be understood as the equilibrium Bose distribution function. In two-dimensional pure systems the momentum integral logarithmically diverges at the infrared. However, in the presence of the elastically scattering potential the dispersion relation of plasmons which we used holds only at those frequencies for which the plasmon mean free time is shorter than the electron-impurity scattering time $\tau$. This determines a cut-off in Eq. \eqref{eq:dkappa}. In Sec. \ref{sec:disorder} we analyze broadening of plasmon dispersion and provide estimates for two different models of the disorder potential. As a result, the fluctuational correction to the thermal conductivity can be estimated as 
$\delta\kappa\sim (e^2/\sigma)E_F\ln(T\tau)$. This estimate should be compared to the thermal conductivity of a two-dimensional electron gas with Coulomb interaction \cite{Lyakhov,PrincipiVignale}:
$\kappa=(E^2_F/T)\ln^{-1}(E_F/T)$. Therefore the correction is small $\delta\kappa/\kappa\sim (T/E_F)\ll1$. 

In complete analogy we can estimate corrections to viscosity. For that one needs to consider the flux of momentum associated with plasmon excitations, which is given by 
\begin{equation}
\Pi_{ij}=-\int\frac{d^2q}{(2\pi)^2} q_i\frac{\partial\omega_q}{\partial q_j}N_{\bm{q}}.
\end{equation}
Solving now the kinetic equation \eqref{eq:KinEq} to the linear order in the hydrodynamic flow with velocity $\bm{u}(\bm{r})$ we extract the correction to the shear viscosity 
\begin{equation}
\delta\eta=-\frac{1}{2}\int q^2\left(\frac{\partial\omega_q}{\partial q}\right)^2\tau_q\frac{\partial N_q}{\partial\omega_q}\frac{d^2q}{(2\pi)^2}, 
\end{equation}
and after integration $\delta\eta\sim(v^2_F/e^2\sigma)n(T/E_F)^3$. This correction is negligible as compared to the electron viscosity in a 2D Fermi liquid, which is $\eta\sim n(E_F/T)^2$ modulo logarithmic factors \cite{Novikov,Alekseev}.  

\section{Plasmon attenuation from disorder scattering}\label{sec:disorder}

For completeness, we briefly recall the effect of disorder on plasmon scattering and relaxation, see Refs. \cite{Principi,Vignale,Briskot,Titov} for the related studies. 
To this end, it suffices to analyze the $\sigma\to0$ limit, namely Galilean invariant case. In the following, we focus on a specific experimentally relevant setup in which a doping layer is separated from the two-dimensional electron system by a distance $d$ \cite{Ando}. Randomness of the spatial distribution of charged dopants leads to electron density in the form 
\begin{equation}
n(\bm{r},t)=n+\widetilde{n}(\bm{r})+\delta n(\bm{r},t)
\end{equation}
where $n$ is the average uniform electron density, $\widetilde{n}(\bm{r})$ is the static density variations due to doping layer, and   
$\delta n(\bm{r},t)$ is the plasmon-related dynamic density oscillations. The average density of dopants is equal to the average density of electrons.  
Within the hydrodynamic approximation density fluctuations follow the equation of motion 
\begin{equation}
\frac{\partial^2}{\partial t^2}\delta n(\bm{r},t)-\frac{e^2}{\epsilon m}\bm{\nabla}[n+\widetilde{n}(\bm{r})]\bm{\nabla}\int d^2\bm{r}'\frac{\delta n(\bm{r}',t)}{|\bm{r}-\bm{r}'|}=0
\end{equation}
In the Fourier representation this equation can be cast in the form 
\begin{equation}
(\omega^2-\omega^2_q)\delta n(\bm{q})=\frac{2\pi e^2q}{\epsilon m}\sum_{\bm{k}}(\mathbf{n}_{\bm{q}}\cdot\mathbf{n}_{\bm{k}})\widetilde{n}(\bm{q}-\bm{k})\delta n(\bm{k})
\end{equation}
where $\mathbf{n}_{\bm{k}}$ is the unit vector along the direction of the corresponding momentum. We solve this integral equation perturbatively performing disorder averaging to the lowest order in $\widetilde{n}$. After one iteration we find 
\begin{equation}\label{eq:plasmon-dis}
\omega^2-\omega^2_q-\Sigma(\omega,q)=0,
\end{equation} 
where the self-energy is given by 
\begin{equation}\label{eq:self-energy}
\Sigma(\omega,q)=\left(\frac{2\pi e^2}{\epsilon m}\right)^2\sum_{\bm{k}}\frac{qk(\mathbf{n}_{\bm{q}}\cdot\mathbf{n}_{\bm{k}})^2}{\omega^2-\omega^2_k+i0}
D(\bm{q}-\bm{k}),
\end{equation}
which is expressed in terms of a correlation function of density variations caused by disorder 
\begin{equation}
D(\bm{q})=\int d^2\bm{r}\langle\widetilde{n}(\bm{r})\widetilde{n}(0)\rangle e^{-i\bm{qr}}.
\end{equation}
In contrast to the notations of the previous sections here $\langle\ldots\rangle$ denotes disorder average rather than thermal average.  
It should be noted that within the theory of linear screening this correlation function is related to the probability of electron scattering since density variations cause potential variations, $\nu\delta V(\bm{r})+\widetilde{n}(\bm{r})=0$, which is Thomas-Fermi condition, and $\nu=m/\pi$ is the density of states. The imaginary part of the self-energy in Eq. \eqref{eq:self-energy}
yields the scattering rate
\begin{align}
\tau^{-1}_q&=-\frac{1}{2\omega_q}\Im\Sigma(\omega_q,q)\nonumber\\
&=\frac{\pi^2e^2q}{2\epsilon mn}\sum_{\bm{k}}(\mathbf{n}_{\bm{q}}\cdot\mathbf{n}_{\bm{k}})^2\delta(\omega_q-\omega_k)D(\bm{q}-\bm{k})
\end{align} 
For the spatially uncorrelated dopants, the spectral power of the external random potential induced in the plane of the
electron system is
\begin{equation}\label{eq:D}
D(q)=\frac{n\varkappa^2}{(q+\varkappa)^2}e^{-2qd},
\end{equation}
where $\varkappa=2\pi e^2\nu$ is the inverse Thomas-Fermi screening radius. For a quantum well of a thickness $a$ the above expression for $D(q)$ should be additionally multiplied by a factor $\sinh(qa)/qa$ \cite{Ando}. In Eq. \eqref{eq:D} we assumed $qa\ll1$. For smooth disorder $(k_F,\varkappa)\ll 1/d$, which leads to a plasmon broadening 
\begin{align}\label{eq:tauq}
\tau^{-1}_q=\frac{\omega_qq^2}{16\pi n}\int^{\pi}_{0}d\theta \cos^2\theta e^{-4qd\sin\frac{\theta}{2}}\nonumber \\
=\frac{q\omega_q}{32\pi nd}\left\{\begin{array}{cc}\pi qd & qd\ll1 \\ 1 & qd\gg1\end{array}\right..
\end{align}

It can be readily verified that the effect of density variation in the random long-range potential on the attenuation of two-dimensional plasmons dominates over the effect of broadening of single-particle states for smooth disorder. Indeed, the latter can be inferred from the optical Drude conductivity  
\begin{equation}
\sigma(\omega)=\frac{ie^2n/m}{\omega+i/\tau_{\text{tr}}}
\end{equation}
expressed in terms of the transport scattering time \cite{Stern}
\begin{equation}
\tau^{-1}_{\mathrm{tr}}=\frac{2\pi}{\nu}\int d\mathbf{n}'(1-\mathbf{n}\cdot\mathbf{n}')D(k_F\mathbf{n}-k_F\mathbf{n}').
\end{equation}
In the hydrodynamic approach \cite{Fetter} the plasmon spectrum is related to the optical conductivity as follows $\omega=-2\pi iq\sigma(\omega)/\epsilon$, which is equivalent to 
\begin{equation}
\omega(\omega+i/\tau_{\text{tr}})=\omega^2_q.
\end{equation}
This equation is analogous to Eq. \eqref{eq:plasmon-dis}. However in contrast, plasmon broadening is independent of $q$ in the limit $\tau^{-1}_{\text{tr}}\ll \omega_q$, and simply determined by the transport scattering rate   
\begin{equation}\label{eq:tautr}
\tau^{-1}_q=\frac{1}{2\tau_{\text{tr}}}=\frac{2\pi n}{\nu}\int^{\pi}_{0}d\theta\sin^2\frac{\theta}{2}e^{-4k_Fd\sin\frac{\theta}{2}}\approx\frac{E_F}{(2k_Fd)^3}.
\end{equation}
It is dominated by a small angle scattering $\theta\sim q/k_F\ll1$. Comparing Eq. \eqref{eq:tauq} and \eqref{eq:tautr} one can see that plasmon broadening from inhomogeneous density variation dominates starting from momenta $q\sim d^{-1}/(k_Fd)^{1/5}<d^{-1}$. In contrast, if we compared now intrinsic Maxwellian rate of attenuation for systems with broken Galilean invariance Eq. \eqref{eq:energies} to that of disorder broadening Eq. \eqref{eq:tauq} at typical moment for bilayers $q\sim d^{-1}$ the former dominates in a large factor $\sim (\sigma/v_F)(k_Fd)^{3/2}\gg1$. These estimates justify assumptions and approximation used in Sec. \ref{sec:drag}.   

\section*{Acknowledgments}

We thank A. Andreev, L. Fritz, L. Glazman, and B. Spivak for discussions that shaped understanding of physics presented in this work. This research project was financially supported by the National Science Foundation Grant No. DMR-2203411 and H. I. Romnes Faculty Fellowship provided by the University of Wisconsin-Madison Office of the Vice Chancellor for Research and Graduate Education with funding from the Wisconsin Alumni Research Foundation. This paper was finalized during the Aspen Center of Physics 2023 summer program on ``Quantum Materials: Experimental Enigmas and Theoretical Challenges", which was supported by the National Science Foundation Grant No. PHY-2210452. 

\bibliography{biblio}

\end{document}